\documentclass[aps,twocolumn,superscriptaddress,showpacs]{revtex4}
\usepackage{amsmath} 
\usepackage{amssymb} 
\usepackage{amsfonts}
\usepackage[dvips]{graphicx} 
\usepackage[]{epsfig} 
\newcommand{\reff}[1]{{\rm (\ref{#1})}}

\begin{document}

\title{Application of the level-set method to the implicit solvation of nonpolar molecules}
\author{Li-Tien Cheng} 
\email[e-mail address:]{lcheng@math.ucsd.edu} 
\affiliation{Department of Mathematics, University of California, San Diego, La Jolla,
California 92093-0112} 
\author{Joachim Dzubiella} 
\email[e-mail address:]
{jdzubiel@ph.tum.de}
\affiliation{Physics Department (T37), Technical University Munich, 
 James-Franck-Stra{\ss}e, 85748 Garching, Germany}
\affiliation{NSF Center for Theoretical Biological Physics (CTBP)} 
\affiliation{Department of Chemistry and Biochemistry, University of California, 
San Diego, La Jolla, California 92093-0365} 
\author{J.~Andrew McCammon} 
\email[e-mail address:]{jmccammon@ucsd.edu} 
\affiliation{NSF Center for Theoretical Biological Physics (CTBP)}
\affiliation{Department of Chemistry and Biochemistry, and Department of Pharmacology, University of
California, San Diego, La Jolla, California 92093-0365}
\author{Bo Li} 
\email[e-mail address:]
{bli@math.ucsd.edu} 
\affiliation{Department of Mathematics, University of California, San Diego, La Jolla,
California 92093-0112} 

\date{\today}

\begin{abstract}
  A level-set method is developed for numerically capturing the
  equilibrium solute-solvent interface that is defined by the recently
  proposed variational implicit solvent model (Dzubiella, Swanson, and
  McCammon, Phys.\ Rev.\ Lett.\ {\bf 104}, 527 (2006) and J.\ Chem.\
  Phys.\ {\bf 124}, 084905 (2006)).  In the level-set method, a
  possible solute-solvent interface is represented by the zero
  level-set (i.e., the zero level surface) of a level-set function and
  is eventually evolved into the equilibrium solute-solvent interface.
  The evolution law is determined by minimization of a solvation free
  energy {\it functional} that couples both the interfacial energy and
  the van der Waals type solute-solvent interaction energy. The
  surface evolution is thus an energy minimizing process, and the
  equilibrium solute-solvent interface is an output of this process.
  The method is implemented and applied to the solvation of nonpolar
  molecules such as two xenon atoms, two parallel paraffin plates,
  helical alkane chains, and a single fullerene $C_{60}$.  The
  level-set solutions show good agreement for the solvation energies
  when compared to available molecular dynamics simulations. In
  particular, the method captures solvent dewetting (nanobubble
  formation) and quantitatively describes the interaction in the
  strongly hydrophobic plate system.
\end{abstract}


\maketitle

\section{Introduction}
\label{s:introduction}

The correct 
%theoretical 
description of solvation free energies and
detailed solution structures of biomolecules is crucial to our
understanding of molecular processes in biological systems.  Efficient
theoretical approaches to such descriptions are typically given by so-called
implicit (or continuum) solvent models of the aqueous environment
\cite{RouxSimonson99,FeigBrooksIII04}.  In those models, the solvent
molecules and ions (e.g., as in physiological electrolyte solutions)
are treated implicitly and their effects are coarse-grained.  In
particular, the description of the solvent is reduced to that of the
continuum solute-solvent interface and related macroscopic quantities,
such as the surface tension and the position-dependent dielectric
constant serving as input or fitting parameters.
%These models are complementary to the more accurate but computationally 
%expensive explicit solvent models which often provide sampled 
%statistical information rather than direct descriptions of thermodynamics.

Most of the existing implicit solvent models are built upon the
concept of solvent accessible surface area (SASA) defined in several
ways \cite{LeeRichards71,Richards77}.  In these models, the solvation
free energy is proportional to the SASA for the nonpolar contribution,
complemented by the Poisson-Boltzmann (PB)
\cite{Fixman79,DavisMcCammon90,SharpHonig90} or Generalized Born (GB)
\cite{StillETC90,BashfordCase00} description of electrostatics, i.e.,
the polar contribution. Although successful in many cases, the general
applicability of these rather empirical models with many
system-dependent, adjustable parameters (e.g., individual atomic
surface tensions) is often questionable, when compared to more
accurate but computationally expensive explicit molecular dynamics
(MD) simulations or experimental results. It is believed that the key
issues here are the decoupling and separate analysis of surface area,
dispersion and polar parts of the free energy, and the inaccurate free
energy estimation due to a predefined solvent-solute interface, an
{\it ad hoc} input. It is additionally well established by now that
cavitation free energies do not scale with surface area for high
curvatures \cite{LCW99,chandler:review}, a fact of critical importance
in the implicit modeling of hydrophobic interactions in biomolecular
systems \cite{brooks}.

Recently, Dzubiella, Swanson, and McCammon \cite{DSM1,DSM2} have
developed a {\it variational} implicit solvent model. The basic idea
of this approach is to introduce a free energy functional of all
possible solute-solvent interfaces, coupling both the nonpolar and
polar contributions of the system, and allowing for curvature
correction of the surface tension to approximate the length-scale
dependence of molecular hydration. Minimizing the functional leads to
a partial differential equation whose solution determines the
equilibrium solute-solvent interface and the minimum free energy of
the solvated system. This stable solute-solvent interface is an output
of the theory. It results automatically from balancing the different
contributions of the free energy.  First applications of this approach
to simple, highly symmetrical solutes showed promising results when
compared to MD simulations \cite{DSM1,DSM2}.

In this work, we develop a {\it level-set method} for numerically
capturing {\it arbitrarily shaped} solute-solvent interfaces
determined by the solvation free energy functional in the variational
implicit solvent model. In our method, a possible solute-solvent
interface is represented by the zero level-set (i.e., the zero level
surface) of a level-set function; and an initial surface is evolved
eventually into an equilibrium solute-solvent interface.  
% The
% evolution law is so determined that the process decreases and
% eventually minimizes the solvation free energy that couples both the
% surface energy and the van der Waals type solute-solvent interaction
% energy.  The surface evolution is thus an energy minimizing process,
% and the equilibrium solute-solvent interface is an output of this
% process.
The level-set method is a general technique for numerically tracking
moving fronts with possible topological changes such as interface
merging and break-ups
\cite{OsherSethian88,Sethian99,OsherFedkiw02}. Previous applications
include two-phase fluid flow, crystal growth, materials modeling,
shape optimization, imaging process and graphics, etc.; see
\cite{Sethian99,OsherFedkiw02} for related references.  Recently, Can,
Chen, and Wang \cite{CanChenWang06} used the level-set method for
imaging a large biomolecular surface based on a SASA type model. Our
work is quite different: we not only represent molecular surfaces
using level-sets, but further develop an evolution algorithm that
numerically determines the stable equilibrium solute-solvent interface
based on physical rationale.

Our new level-set
techniques include two efficient and stable methods. One is a careful
treatment of singularities formed during the level-set evolution based
on certain geometrical motion of the molecular surface.  The other is
a two-grid numerical method for the calculation of the free energy
using a Lennard-Jones type potential which changes dramatically for
short range interactions.

We apply our method to the solvation of nonpolar molecules such as two
xenon atoms, two paraffin plates, model helical alkane chains, and a
single $C_{60}$ molecule.  Our extensive numerical results show good
agreement with MD calculations. In particular, our method is able to
capture the solvent dewetting phenomenon, i.e., formation of a
nanobubble within the strong hydrophobic confinement caused by the
paraffin plate arrangement. Furthermore, we demonstrate that
topological changes, such as the rupture of a nanobubble and the
fusion/breakup of the surface of two molecules, are captured by our
level-set method.

The rest of the paper is organized as follows: In
Section~\ref{s:vism}, we review the variational implicit solvent
model. This is followed by a description of our level-set method in
Section~\ref{s:levelset}. In Section~\ref{s:applications}, we report
the numerical results of our level-set calculation of several nonpolar
systems. Finally, in Section~\ref{s:conclusions}, we conclude and give
an outlook to further necessary extensions of our approach.

\section{Variational implicit solvent approach}
\label{s:vism}

In the following, we briefly summarize, with a few remarks, the
variational implicit solvent approach that has been recently proposed
by Dzubiella, Swanson, and McCammon \cite{DSM1,DSM2}.

\subsection{Geometry}

Consider the system of an assembly of solutes with arbitrary shape and
composition surrounded by a solvent. Let us denote by $\cal W$ the
region of the entire system that includes both the solute and solvent
regions.  Let us  denote by $\cal V$ the region of solutes, or the
cavity region, which is empty of solvent.  We identify the {\it
  solute-solvent interface} to be the boundary of the solute region
$\cal V$, and denote it by $\Gamma = \partial \cal V$.  We assume that
the surface $\Gamma$ consists of possibly many connected components,
each of which is closed and continuous.

For the cavity region $\cal V$, we assign a volume exclusion function
\[
v(\vec r)= \begin{cases} 0 & {\text {for}} \,\,\vec r \in \cal V, \cr 1 
& {\rm else}. \cr
\end{cases}
\]
Mathematically, this is the characteristic function of the solvent
region $\cal W \setminus \cal V$ which is the set of points in $\cal
W$ but not in $\cal V$.  The volume $\mbox{Vol}\,[\cal V]$ of the
solute region $\cal V$ and the surface area $\mbox{Area}\,[\Gamma]$ of
the interface $\Gamma$ can then be expressed as functionals of
the volume exclusion function $v(\vec r)$ via
\begin{eqnarray}
\mbox{Vol}\,[{\cal V}]=\int_{\cal V} d^3r =  
\int_{\cal W} [1-v(\vec r)]\, {\rm d}^3r, \nonumber \\
\mbox{Area}\,[\Gamma]
= \int_\Gamma dS = \int_{\cal W} |\nabla v(\vec r)| \, {\rm d}^3r, \nonumber 
\end{eqnarray}
where $\nabla\equiv\nabla_{\vec r}$ is the usual gradient operator
with respect to the position vector $\vec r$ and $|\nabla v(\vec r)|$
is the $\delta$-function concentrated on the boundary $\Gamma
= \partial \cal V$ of the cavity region $\cal V$.  The expression $dS
\equiv |\nabla v(\vec r)| \, {\rm d}^3r $ can thus be identified as
the infinitesimal surface element. We remark that, within the framework
of the variational implicit solvent model \cite{DSM1,DSM2}, either the
volume exclusion function $v$ of the cavity or its boundary $\Gamma$
can be used as the ultimate, direct variable of the solvation free
energy of an underlying system.

We assume that the position of each solute atom ${\vec r}_i$ and the
solute conformation are fixed. Thus, the solutes can be considered as
an external potential to the solvent without any degrees of freedom.
In this continuum solvent model, the solvent density distribution is
simply $\rho(\vec r)=\rho_0 v(\vec r)$, where $\rho_0$ is the bulk
density of the solvent. This means that we use a sharp interface
approximation.

%%%%%%%%%%%%%%%%%%%%%%%%%%%%%%%%%%%%%%%%%%%%%%%%%%%%%%%%%%%%
\subsection{Free energy functional}

For a given solvation system characterized by the cavity region $\cal
V$ with its boundary $\Gamma$, the solute-solvent interface, the
following ansatz of the Gibbs free energy was proposed in
\cite{DSM1,DSM2} as a functional of the volume exclusion function
$v(\vec r)$ or its boundary $\Gamma$:
%\begin{eqnarray}
%\label{eq:grand}
%&G[v]&=G_{\rm pr}[v]+ G_{\rm int}[v]+ G_{\rm ne}[v]\\
%&=& P V[v]+\int_{\cal W}{\rm d}^3r\;\gamma(\vec r;[v])|\nabla v(\vec r)|\nonumber \\
%&+&\rho_0\int_{\cal W}{\rm d}^3r\;v(\vec r) U(\vec r)\nonumber
%\end{eqnarray}
\begin{align}
\label{eq:grand}
G[v]&=G_{\rm vol}[v]+ G_{\rm sur}[v]+ G_{\rm vdW}[v] + G_{\rm ele}[v] \nonumber \\
&= P \int_{\cal W} [1-v(\vec r)]\, {\rm d}^3r +
\int_{\cal W} \gamma(\vec r, \Gamma) | \nabla v (\vec r)|\, d^3{\rm r} \nonumber \\
&\quad +\rho_0\int_{\cal W}
U(\vec r) v(\vec r) \, {\rm d}^3r + G_{\rm ele}[v]. 
\end{align}

The first term, 
\begin{equation}
\label{term1}
G_{\rm vol}[v] = P \int_{\cal W} [1-v(\vec r)]\, {\rm d}^3r   
% = P \int_{\cal V} \, {\rm d}^3 r 
= P \, \mbox{Vol}\, [{\cal V}], 
\end{equation}
proportional to the volume of $\cal V$, is the energy of creating a
cavity in the solvent against the difference in bulk {\it pressure}
$P$ between the liquid and vapor phase, $P=P_l-P_v$.  In water at
normal conditions or any fluid close to phase coexistence, this
pressure difference is small and can often be neglected for
solutes of microscopic size ($\sim$nm).

The second term 
\begin{equation}
\label{term2}
G_{\rm sur}[v] = \int_{\cal W} \gamma(\vec r, \Gamma) | \nabla v (\vec r)|\, d^3{\rm r} 
= \int_{\Gamma}\gamma(\vec r,\Gamma) \,dS 
\end{equation}
describes the energetic cost due to the solvent rearrangement around
the cavity, i.e., near the solute-solvent interface $\Gamma$,
% with area $\mbox{Area}\,[\Gamma]$
in terms of a function $\gamma(\vec r,\Gamma)$ with dimensions of free
energy/surface area.
% that is defined on the interface $\Gamma$.
This surface energy penalty is thought to be the main driving force
behind hydrophobic phenomena \cite{chandler:review}.  It is a solvent
specific quantity that also depends on the particular topology of the
solute-solvent interface and varies locally in space \cite{zwanzig}.

The exact form of $\gamma(\vec r, \Gamma)$ is not known.  The
following approximation based on a first-order curvature correction
from scaled-particle theory \cite{spt} was made in \cite{DSM1,DSM2}:
% in which $\gamma(\vec r, \Gamma )$ reduces to the function
\begin{equation}
\label{H}
\gamma(\vec r, \Gamma )=\gamma_0 [ 1-2\tau H(\vec r)],
\end{equation}
where $\gamma_0$ is the constant solvent liquid-vapor surface tension
for a planar interface, $\tau $ is a positive constant often called
the {\it Tolman length} \cite{Tolman49}, and
\[
H(\vec r)=\frac12 \left[ \kappa_1(\vec
r)+\kappa_2(\vec r) \right]
\]
 is the local mean curvature in which
$\kappa_1(\vec r)$ and $\kappa_2(\vec r)$ are the local principal
curvatures of the interface $\Gamma$. 
% \cite{prince}. 

It has been shown in MD simulations that the surface tension
$\gamma_0$ is the asymptotic value of the solvation free energy per
unit surface area for hard spherical cavities in water in the limit of
large radii \cite{HuangGeisslerChandler01,LCW99}. In this system the
Tolman length has been estimated to be of molecular size and has
values of 0.7--0.9\AA.  As its exact value is not known, the Tolman
length may serve as the only fitting parameter in the variational
continuum solvent model. The mean curvature $H$ is defined only on the
solute-solvent interface $\Gamma$.  We have chosen the convention in
which the curvatures are positive for convex surfaces (e.g., a
spherical cavity) and negative for concave surfaces (e.g., a spherical
droplet).

We remark that, by the Hadwiger Theorem in differential geometry
\cite{Hadwiger57}, the geometrical part of the energy as a valuation
of the closed surface $\Gamma$ should have all the terms in $G_{\rm vol} +
G_{\rm sur}$ (volume, surface area, and surface integral of the mean
curvature) plus an additional term of the surface integral of the
Gaussian curvature
\[
K(\vec r)=\kappa_1(\vec r)\kappa_2(\vec r). 
\]
But, by the Gauss-Bonnet Theorem \cite{Kreyszig91}, the Gaussian
curvature is an intrinsic geometric property of the surface $\Gamma$,
and its contribution to the free energy is an additive constant.
Therefore, it does not change our energy minimization process. We note
that the Hadwiger Theorem was used in a generalization of the
classical theory of capillarity \cite{BoruvkaNeumann77}, and 
recently in a morphometric approach to solvation
\cite{KonigRothMecke04, RothETC06}.

The third term
\begin{equation}
\label{term3}
G_{\rm vdW}[v]
= \rho_0\int_{\cal W} U(\vec r) v(\vec r)  \, {\rm d}^3r 
= \rho_0\int_{{\cal W}\setminus {\cal V}} U(\vec r) {\rm d}^3r
\end{equation}
is the total energy of the non-electrostatic, {\it van der Waals}
type, solute-solvent interaction given a solvent density distribution
$\rho_0v(\vec r)$. The potential
\begin{equation}
\label{UUi}
U(\vec r)=\sum_{i=1}^N U_i(|\vec r-\vec r_i|)
\end{equation}
is the sum of $U_i$ that describes the interaction of the $i$th solute
atom (with $N$ total atoms) centered at $\vec r_i$ with the
surrounding solvent.  Each term $U_i$ includes the short-ranged
repulsive exclusion and the long-ranged attractive dispersion
interaction between each solute atom $i$ at position $\vec r_i$ and a
solvent molecule at $\vec r$.
%The integral in $G_{\rm vdW}[v]$ is over the solvent region $\cal W \setminus \cal V$. 
Classical solvation studies typically represent the
interaction $U_i$ as an isotropic Lennard-Jones (LJ) potential,
\begin{equation}
\label{ULJ}
U_{\rm LJ}(r)=4\epsilon
\left[\left(\frac{\sigma}{r}\right)^{12}-\left(\frac{\sigma}{r}\right)^6\right],
\end{equation}
with an energy scale $\epsilon$, length scale $\sigma$, and
center-to-center distance $r$.

The last term $G_{\rm ele}[v]$ is the electrostatic energy due to
charges possibly carried by solute atoms and the ions in the solvent.
In this work, we only consider nonpolar solutes. Therefore, we shall
neglect this term in what follows and refer to
\cite{DavisMcCammon90,DSM2,Sharp95,SharpHonig90} and our forthcoming
work \cite{CCDLM07} for details. With this and
considerations~\reff{eq:grand}--\reff{UUi}, we find for the final form
of the nonpolar free energy functional
\begin{align}
\label{eq:grandAgain}
G[v] &
= P \, \mbox{Vol}\, [{\cal V}] + 
 \int_{\Gamma} \gamma_0 [ 1-2\tau H(\vec r)] \,dS \nonumber\\ 
&\quad + \rho_0\sum_{i=1}^N \int_{{\cal W}\setminus {\cal V}} 
U_{i} (|\vec r-\vec r_i|) {\rm d}^3r, 
\end{align}
where each interaction potential $U_i$ $(1 \le i \le N)$ has the 
form \reff{ULJ}. 
%is the sum $\sum_{i}$ is taken over all the fixed solute atoms centered at $\vec r_i$.  
%This is the free energy functional we use in this work. 

%Add Hadwiger Theorem, 
%%%%%%%%%%%%%%%%%%%%%%%%%%%%%%%%%%%%%%%%%%%%%%%%%%%%%%%%%%%%%%%%%%%%%%%%%%%%
\subsection{Free energy minimization}

Let $v_{\rm min}(\vec r)$ with its boundary $\Gamma_{\rm min}$ be the
exclusion function which minimizes the functional
\reff{eq:grandAgain}.  Then, the resulting Gibbs free energy of the
system is given by $G[v_{\rm min}]$. The solvation free energy $\Delta
G$ is the reversible work to solvate the solute and is given by
\[
\Delta G=G[v_{\rm min}]-G_0,
\]
where $G_0$ is a constant reference energy which can refer to the pure
solvent state and an unsolvated solute.  The solvent-mediated
potential of mean force along a given reaction coordinate $x$ (e.g.,
the distance between two solute centers of mass) is given, up to an
additive constant, by $w(x)=G[v_{\rm min}]$, where $v_{\rm min}(\vec
r)$ must be evaluated for every $x$.

A necessary condition for $\Gamma$ to be an energy minimizing
solute-solvent interface is that the first variation of the free energy
functional \reff{eq:grandAgain} vanishes at the corresponding 
 volume exclusion function $v$, i.e., 
\begin{equation}
\label{dGdv0}
%\delta G[v]/\delta v= 0
\frac{\delta G[v]}{\delta v}= 0
\end{equation}
at every point of the boundary $\Gamma$. 
%where taken the free energy functional \reff{eq:grandAgain} 
This energy variation can be identified as a
distribution over the interface $\Gamma$, and is given by
\cite{DSM1,DSM2}
%\begin{eqnarray}
\begin{equation}
\label{dGdv}
\frac{\delta G[v]}{\delta v} 
 =  P+2\gamma_{{0 }}\left[H(\vec r)-\tau K(\vec r)\right]-\rho_0 U(\vec r). 
%\frac{\delta G[v]}{\delta v} & = & 0 \cr
% & = & P+2\gamma_{{0 }}\left[H(\vec r)-\tau K(\vec r)\right]-\rho_0 U(\vec r). 
%\end{eqnarray}
\end{equation}
% In deriving this formula, we utilize the following two formulas
% \[
% \frac{\delta}{\delta v}\int_{\cal W} |\nabla v(\vec r)|\, {\rm d}^3r 
% =\frac{\delta}{\delta v}\int_{\Gamma}{\rm d}S=2H
% \]
% and
% \[
% \frac{\delta}{\delta v}\int_{\cal W} H(\vec r) |\nabla v(\vec r)|\, {\rm d}^3r
% =\frac{\delta}{\delta v}\int_{\Gamma} H(\vec r)\, {\rm d}S=K. 
% \]
% These formulas have been derived in detail by Helfrich and Ouyang by 
% means of differential geometry \cite{helfrich,helfrich2}.
 The partial differential
equation (PDE) determined by \reff{dGdv0} and \reff{dGdv} 
for the optimal exclusion function $v_{\rm min}(\vec
r)$, or equivalently the optimal solute-solvent interface $\Gamma_{\rm
  min}$, is expressed in terms of pressure, curvatures, short-range
repulsion, and dispersion, all of which have dimensions of energy
density.  It can  be interpreted as a mechanical balance between
the forces per surface area generated by each of the particular
contributions.  A similar expression without the dispersion term was
derived by Boruvka and Neumann within a generalization of
classical capillarity \cite{BoruvkaNeumann77}.

% If the curvature correction ($K$-term) and the dispersion term are neglected, 
% one then obtains the classical Laplace-Young equation 
% \[
% P=-2\gamma_{0}H,
% \] 
% which is exclusively used for the shape description of macroscopic
% capillary and interfacial phenomena in conjunction with appropriate
% boundary conditions, e.g., prescribed liquid-solid contact angles at
% the solid surfaces \cite{kralchevsky,safran}.

%%%%%%%%%%%%%%%%%%%%%%%%%%%%%%%%%%%%%%%%%%%%%%%%%%%%%%%%%%%%%%%%%%%%%%%%%%%%%%%%
\section{The level-set method}
\label{s:levelset}

%%%%%%%%%%%%%%%%%%%%%%%%%%%%%%%%%%%%%%%%%%%%%%%%%%%%%%%%%%%%%%%%%%%%%%%%%%%%%%%%
\subsection{Basics}

The starting point of the level-set method is to identify a surface
$\Gamma $ in  three-dimensional space as the zero level-set of a
function $\phi = \phi(\vec{r})$
\cite{OsherSethian88,OsherFedkiw02,Sethian99}:
\[
 \Gamma = \{ {\vec r} :  \phi (\vec r) = 0 \}.  
\]
This means that the surface consists exactly of those points $\vec r$
at which the function $\phi$ vanishes.  This is in contrast with a
parametric description of the surface $\Gamma: {\vec r} = {\vec r}
(\alpha,\beta)$ with $\alpha, \beta $  the parameters. The function $\phi = \phi (\vec r)$
is called a {\it level-set function} of the surface $\Gamma$.
%Figure 3.5 shows a moving curve (a two-dimensional ``surface'') 
%represented by the zero level-set of a level-set function, where  
Clearly, the level-set function whose zero level-set represents 
the surface $\Gamma$ is vastly non unique.

The level-set function $\phi$ can be used to calculate many important
geometrical quantities of the surface $\Gamma$. For instance, the unit
normal vector $\vec n$ at the interface $\Gamma$, the mean curvature
$H$, and the Gaussian curvature $K$ can all be expressed in terms of
the level-set function $\phi$:
\begin{equation}
\label{nHK}
\vec n = \frac{\nabla \phi}{| \nabla \phi |}, \quad
H = \frac12 \nabla \cdot \vec n, \quad
K = \vec n \cdot \mbox{adj}\,(He ( \phi )) \vec n, 
\end{equation}
where $He( \phi)$ is the $3 \times 3$ Hessian matrix of the function
$\phi$ whose entries are all the second order partial derivatives
$\partial^2_{ij} \phi$ of the level-set function $\phi$, and
$\mbox{adj}\, (He(\phi)) $ is the adjoint matrix of the Hessian
$He(\phi)$.

Consider now a moving surface $\Gamma = \Gamma(t)$ at time $t$.  Let
$\phi = \phi (\vec r, t)$ be a level-set function that represents the
surface $\Gamma(t)$ at time $t$.  The basic idea is now to track the
motion of the moving surface $\Gamma(t)$ by evolving the level-set 
function $\phi(\vec r, t)$ and its zero level-set 
%locating the zero level-set of the level-set function $\phi = \phi (\vec r, t) $ at each
at each time $t$.  The level-set function is determined by the so-called {\it
  level-set equation},
\begin{equation}
\label{lse}
\partial_t \phi + v_n | \nabla \phi | = 0, 
\end{equation}
where $v_n$ is the normal velocity at the point 
$\vec r$  on the surface $\Gamma(t)$. 
% and $\vec n$  refers to the unit normal of the surface at that point.  
This normal velocity $v_n = v_n (\vec r (t))$ of each point $\vec r =
\vec r(t)$ on the surface $\Gamma = \Gamma(t)$ at time $t$ is defined
by
\begin{equation}
\label{definevn}
v_n = v_n (\vec r (t)) = \frac{d \vec r(t)}{dt} \cdot \vec n. 
\end{equation}
%where $\vec n$ is the unit normal along the surface $\Gamma(t)$ at the point $\vec r(t)$. 
%Note that in general the level-set velocity does not necessarily have units of a
%physical velocity (length/time), as the level-set evolution may not mirror
%the real dynamics of the system. 
%In applications, however, the level-set normal
%velocity  is typically defined from the underlying laws of
%geometry and/or physics and might have units of pressure (see below)
%or force, etc.  
The velocity is usually extended away from the surface
so that the level-set equation \reff{lse} can be solved in a finite
computational box.

%\begin{figure}[htb]
%\psscalefirst
%\psrotatefirst
%\centerline{\psfig{figure=levelset.ps,height=1in,width=2.8in}}
%\centerline{\small Figure~4.1.}
%\end{figure}

One of the major advantages of the level-set method is its easy
handling of topological changes of surfaces during the surface
evolution. For instance, the merge or break of bubbles can be captured
by level-set calculations.  This method is thus a perfect choice for
capturing different kinds of stable solute-solvent interface
structures.

%%%%%%%%%%%%%%%%%%%%%%%%%%%%%%%%%%%%%%%%%%%%%%%%%%%%%%%%%%%%%%%%%%
\subsection{Normal velocity}

We apply the level-set method to evolve an initial interface to an
equilibrium solute-solvent interface. This means that our level-set
evolution is an optimization process rather than the real dynamics of
the solvation system.  We need to choose the governing normal velocity
of the interface in such a way that it will decrease the free energy
during the surface evolution.  As in common practice, we define the
normal velocity of level-set to be the negative of the first variation of
the Gibbs free energy:
\begin{equation}
\label{vnvn}
v_n = - \frac{\delta G[v]}{\delta v}.  
\end{equation}
By \reff{dGdv}, the normal velocity is a function defined
on $\Gamma$, and is given by 
\begin{equation}
\label{expvn}
v_n = - P- 2\gamma_{{0 }}\left[H(\vec r)-\tau 
K(\vec r)\right]+\rho_0 U(\vec r). 
\end{equation}
Here, we choose the unit normal $\vec n$ at $\Gamma$ to point from the
solute to the solvent region. 
%The level-set velocity in our method has
%units of pressure which is equal to energy per unit volume or force per unit surface area.

Notice that the interface $\Gamma = \Gamma(t)$, the volume exclusion
function $v= v(t)$, and the normal velocity $v_n = v_n(t)$ all depend
on the time $t$. This is not the time in the real dynamics but rather
represents the state of optimization iteration.  
In particular, the normal velocity does not represent that of the 
interface evolution in real dynamics of the system, and hence
can have non-physical units. 
%Note that in general the level-set velocity does not necessarily have units of a
%physical velocity (length/time), as the level-set evolution may not mirror
%the real dynamics of the system. 
It follows from the
Chain Rule and the definition of the normal velocity \reff{definevn}
that the time derivative of the Gibbs free energy $G[v(t)]$ is
\begin{align*}
\frac{d}{dt} G[v(t)] & =\int_{\Gamma(t)}  \frac{\delta G[v(t)]}{\delta v} 
\left[ 
\frac{d \vec r(t)}{dt} \cdot \vec n \right] dS \\
& = - \int_{\Gamma(t)} \left| \frac{\delta G[v(t)]}{\delta v } \right|^2 d S 
 \le 0. 
\end{align*}
This shows that the normal velocity defined by \reff{vnvn}
decreases the energy. 

With a given initial surface, we solve the level-set equation
\reff{lse} each time step with the normal velocity given by
\reff{expvn} until a steady-state is reached. The steady-state
solution gives a stable, energy minimizing, solute-solvent interface.

%%%%%%%%%%%%%%%%%%%%%%%%%%%%%%%%%%%%%%%%%%%%%%%%%%%%%%%%%%%%%%%%%%%%%%%%%
\subsection{Implementation}

Our level-set algorithm consists of the following steps: 
%Here is a brief description of our level-set algorithm. 

{\it Step 1.} Input parameters and initialize an interface.  The
physical parameters include the pressure difference $P$, the
macroscopic surface tension $\gamma_0$, the Tolman length $\tau$, the
water density $\rho_0$, the LJ energy and length parameters
$\epsilon_i$ and $\sigma_i$, and the coordinates of centers $\vec r_i$
of all the fixed solute atoms $i$. An initial interface is defined by
its level-set function;

{\it Step 2.} Calculate the unit normal $\vec n$, the mean curvature
$H$, and the Gaussian curvature $K$ by \reff{nHK}. Calculate the
free energy using the formula \reff{eq:grandAgain}; 

{\it Step 3.} Calculate the normal velocity
$v_n$ using the formula \reff{expvn},
and extend the normal velocity $v_n$ to the entire computational domain; 

 {\it Step 4.} 
Solve the level-set equation \reff{lse}. As usual, this equation
needs to be solved only locally near the interface $\Gamma$, since
the value of $\phi$ away from $\Gamma$ will not affect the location
of $\Gamma$; 

{\it Step 5.}  Reinitialize the level-set function $\phi$.  The
gradient of a solution to the level-set equation \reff{lse} at certain
time $t$ can be sometimes too large or too small. This can lead to an
inaccurate approximation of the normal $\vec n$, the mean curvature
$H$, and the Gaussian curvature $K$ by \reff{nHK}.  The
reinitialization process uses the solution of the level-set equation
\reff{lse} to obtain a new level-set function $\phi$ that has the same
zero level-set, i.e., the location $\Gamma$ is not changed, and that
keeps the gradient $|\nabla \phi | $ away from $0$ or from being too
large;

{\it Step 6.} 
Locate the interface $\Gamma$ by the level-set function obtained in the
previous step.  Update the time step and go back to {\it Step 2}. 

%We address two  numerical issues in our level-set calculations. 
%We have developed stable numerical methods in the implementation of our
%level-set method for determining solute-solvent interfaces. 
There are two sources of instability in our level-set calculations.
One is the Gaussian curvature term in the normal velocity
\reff{expvn}.  Elementary calculations show that the motion by the
combination of the mean and Gaussian curvatures, the $H$ and $K$ terms
in \reff{expvn}, results in a parabolic equation of degenerate type in
certain parameter regimes. Specifically, one or two of the three 
eigenvalues of a matrix that defines the type of differential
equations can become $0$ or even negative.  In such a case, we add a small, positive
constant to such eigenvalues so that the evolution is
regularized. Such a regularization does not affect the final
equilibrium solution.

The other is the rapid change of values of the Lennard-Jones potential
\reff{ULJ} when the distance is small.  This can easily lead to large
errors in the calculation of the free energy.  To deal with this
instability, we have developed a two-grid algorithm.  Our idea is to
evolve the level-set function on a coarse grid and to calculate the
energy on a fine grid. In doing so, we use interpolation and
projection techniques to pass along information between the two
grids. Numerical results show that this treatment works very well.
%Our initial calculations show that this is an efficient and accurate 
%way to calculate the energy. We will further improve this hybrid discretization 
%algorithm, and seek other efficient energy calculation techniques as well. 

%%%%%%%%%%%%%%%%%%%%%%%%%%%%%%%%%%%%%%%%%%%%%%%%%%%%%%%%%%%%%%%%%%%%%%%%%
\section{Applications}
\label{s:applications}

In this section, we report on our level-set calculations of four
nonpolar systems and compare the results to available MD
simulations using the SPC/E water model \cite{berendsen:jpc}. The four
systems are two xenon atoms, two helical alkane chains, a single
fullerene $C_{60}$, and two paraffin plates.

In all of our calculations, we focus on water close to normal
conditions ($T=298$K and $P$=1bar) so that the pressure term in the
free energy can be neglected.  The other parameters are the
macroscopic surface tension $\gamma_0$, the Tolman length $\tau$, and
the water density $\rho_0$. The surface tension for SPC/E water has
been calculated to be $\gamma_0=72$mN/m in agreement with the value of
real water \cite{alejandre}. The density is
$\rho_0=0.033$\AA$^{-3}$. The Tolman length has been estimated roughly
for SPC/E to be $\tau=0.9$\AA~\cite{HuangGeisslerChandler01} and will
serve as our only fitting parameter. All solutes are modeled by
assemblies of identical and uncharged spheres interacting with the LJ
potential (\ref{ULJ}) with energy and length parameters $\epsilon$ and
$\sigma$ as summarized in Tab.~I. These solutes are assumed to be in
fixed configurations and have no degrees of freedom.

\begin{table}
\begin{center}
\begin{tabular}{l |   c c }
  System  & $\epsilon$/$k_BT$ & $\sigma$/\AA   \\ 
  \hline 
  Xenon   &  0.431 & 3.57  \\
  Helix  (CH$_2)$  &  0.265 & 3.54     \\
  C$_{60}$ (C)&  0.158 & 3.19 \\
  Plate (CH$_2)$   &  0.265 &  3.54 \\
\end{tabular}
\caption{Investigated system LJ parameters: the atom-water LJ energy
  parameter $\epsilon$ is in units of the thermal energy $k_BT$, and
  the atom-water LJ length $\sigma$ is in \AA.}
\label{tab}
\end{center}
\end{table}

For the helical alkanes and the C$_{60}$, we perform additional MD
simulations to provide data for the solvation free energy that is not
available in literature.  The simulations are done using the MD
simulation package DLPOLY \cite{dlpoly} in the NPT ensemble with up to
$N=800$ SPC/E water molecules. The solvation free energy is calculated
using standard thermodynamic integration \cite{AllenTildesley87}
procedures. Here, at least 15 simulation runs for different
integration parameters $\lambda\in[0,1]$ per system are considered
with 100ps equilibration time and 1ns for gathering statistics, where
$\lambda$ corresponds to the scaled LJ length. The obtained
ensemble-averages were interpolated by an Akima spline and the
resulting curve integrated to get the solvation free energy.

In the level-set method, we usually start with a large spherical
surface in a cubic box that encloses all the fixed solute atoms and
then evolve the surface to minimize the nonpolar solvation energy.
The box length is between 15-25\AA~ and a grid size of $50^3$ or
$100^3$ bins is used depending on the solute size and desired
computational speed and accuracy. Finite-size corrections to the
dispersion part of the energy are considered by integrating the
long-ranged LJ interactions over a homogeneous water distribution
beyond the box up to infinity. The computational time of a single
level-set minimization takes several hours on a single-processor
computer but significantly depends on the choice of the initial
configuration. Independent of the latter, we find that the system
always converges to a stable configuration corresponding to an energy
minimum. We have not investigated whether the systems exhibit multiple
energy minima and postpone this rather complex issue to a future
study.

%%%%%%%%%%%%%%%%%%%%%%%%%%%%%%%%%%%%%%%%%%%%%%%%%%%%%%%%%%%%%%%%%%%%%%%%%
%\subsection{PMF between two noble gas atoms}
\subsection{Two xenon atoms}
\label{ss:twoatoms}

In this example, we investigate the performance of our method for a
simple system of {\it microscopic} nonpolar solutes that consists of
two xenon (Xe) atoms.  For that, we fix the Xe atoms in a
center-to-center distance $d$ and calculate the solvent-mediated
interaction $w(d)$ between them, which is basically the solvation
energy in dependence of the coordinate $d$. The total potential of
mean force (pmf) $W(d)$ is the sum of the solvent-mediated part and
the intrinsic Xe-Xe vacuum interaction which can be modeled also by
an LJ interaction. The water-xenon LJ interactions we use for our
level-set calculations are taken from Paschek \cite{Paschek04} who
accurately calculated the pmf in MD simulations. For our comparison,
we find that Tolman lengths between $\tau=0.9$ and 1.0\AA~ give
reasonable agreement when compared to the MD results, close to the
value of 0.9\AA~estimated by previous MD simulations.

\begin{figure}[bhpt]
\centerline{\psfig{figure=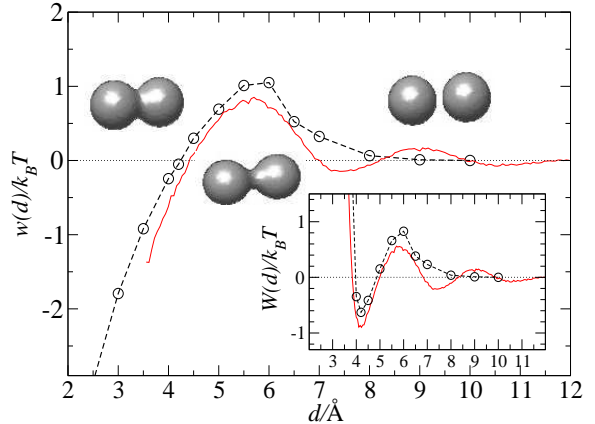,height=2.2in,width=3in}}
\label{f:xenon}
\caption{Comparison of the interaction between two xenon atoms from
  level-set and MD calculations.  The solid line and dotted line with
  symbols (circles) are the solvent-mediated interaction $w(d)$ by the
  MD and level-set calculations, respectively.  The inset displays the
  full pmf $W(d)$ where the vacuum Xe-Xe LJ interaction is added to
  the solvent-mediated part.}
\end{figure}

In Fig.~1, we plot the solvent-mediated part $w(d)$ and the total pmf
$W(d)$ of our level-set calculation compared to the MD simulation
reported by Paschek \cite{Paschek04} for a Tolman length
$\tau=0.95$\AA.  Also plotted are interface images from the level-set
solution at three selected distances. For small separations $d\lesssim
4.5$\AA, the solution gives two overlapping spheres and the
solvent-mediated interaction is attractive in agreement with the MD
simulations. The attraction comes from a smaller water-accessible
surface due to the overlap of spheres and the gain of resulting
interfacial energy.  At separations $d\simeq 5.5$\AA, a maximum of
about 1$k_BT$ in height occurs in both MD and level-set calculations,
also in good agreement with each other. In the continuum approach,
this desolvation barrier is implicitly accounted for by the
unfavorable LJ interaction when replacing water molecules adjacent to
the first Xe atom by the second Xe atom. For separations $d\gtrsim
6$\AA, the interface breaks (water penetrates) and the stable
level-set solution corresponds to two isolated spheres.  The shallow
oscillations in the MD curve for $d\gtrsim 7.5$\AA~ are due to
explicit water structuring around the Xe atoms and are not captured by
our continuum method. Overall, however, we can judge that the
agreement is good and the dominant features of the pmf are well
described by our macroscopic method using just one fit-parameter which
is close to previous estimates.

%%%%%%%%%%%%%%%%%%%%%%%%%%%%%%%%%%%%%%%%%%%%%%%%%%%%%%%%%%%%%%%%%%%%%%%%%%%%%%%%
\subsection{Helical alkane chains}
\label{ss:helix}

In order to check how our level-set method performs for larger and
more complex shaped molecules, we study model helical alkane chains
that are assembled by CH$_2$ atoms modeled by the OPLS-AA force field
\cite{RizzoJorgensen99}; see Tab.~I. We investigate the solvation of
two different configurations, one more loosely packed involving 32
atoms (alkane A) and the other one more tightly packed using 22 atoms
with hardly room for water in the helical core (alkane B).

The results are plotted in Fig.~2 which include a comparison of our
level-set calculation and a typical SASA type surface constructed by
taking the envelope of all the LJ spheres.  Though they look quite
similar, our level-set result leads to a much smoother solute-solvent
interface, a result from the minimization of interface area based on
the energy functional.
 
\begin{figure}[hbtp]
\centerline{\psfig{figure=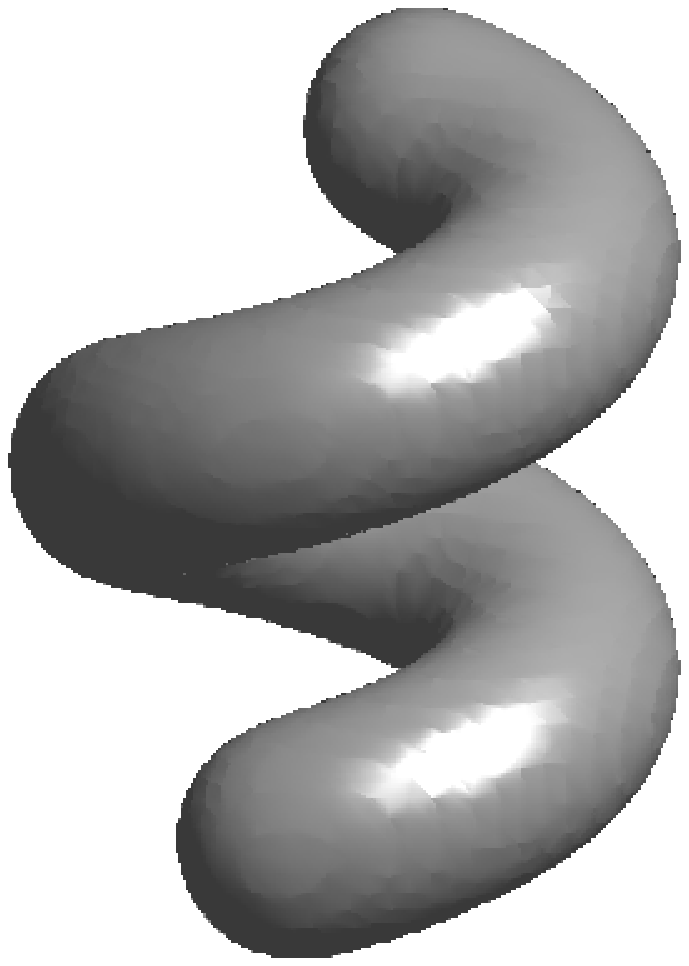,height=1.5in,width=1.5in}
\psfig{figure=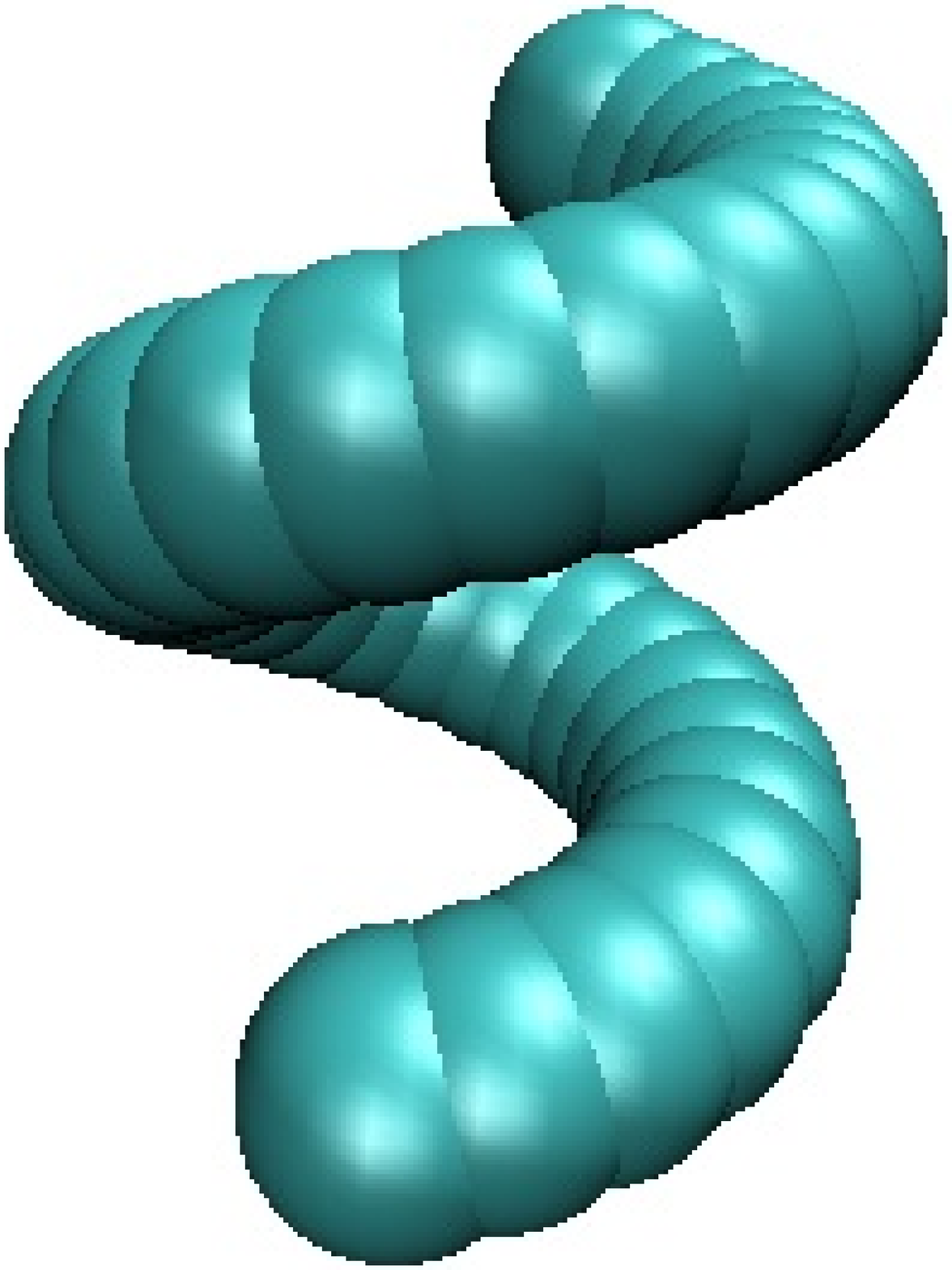,height=1.2in,width=1in} }
\centerline{\psfig{figure=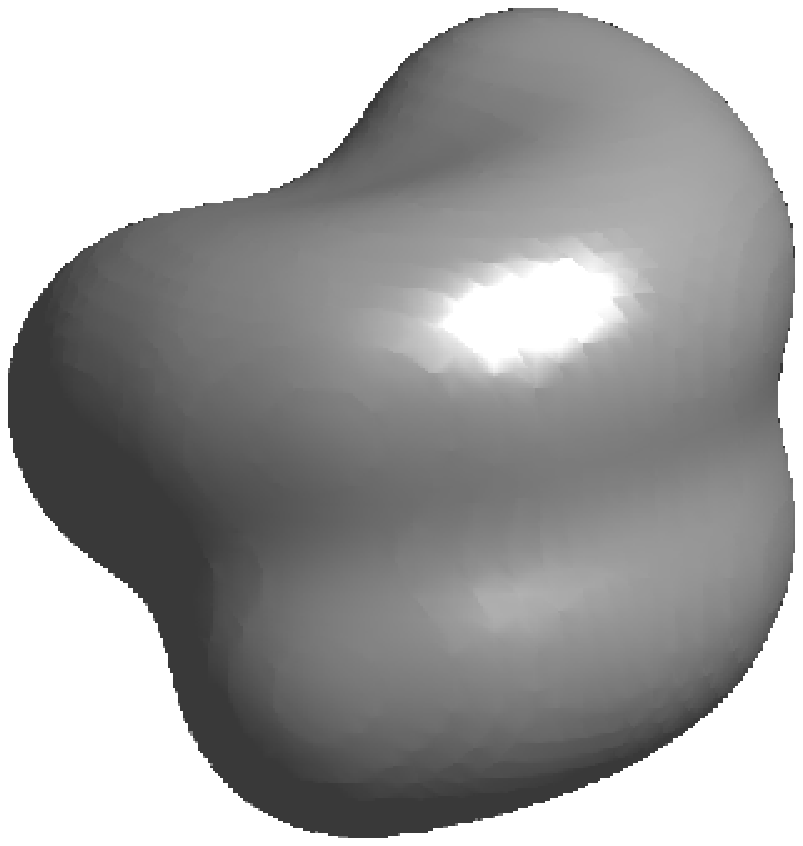,height=1.5in,width=1.5in}
\psfig{figure=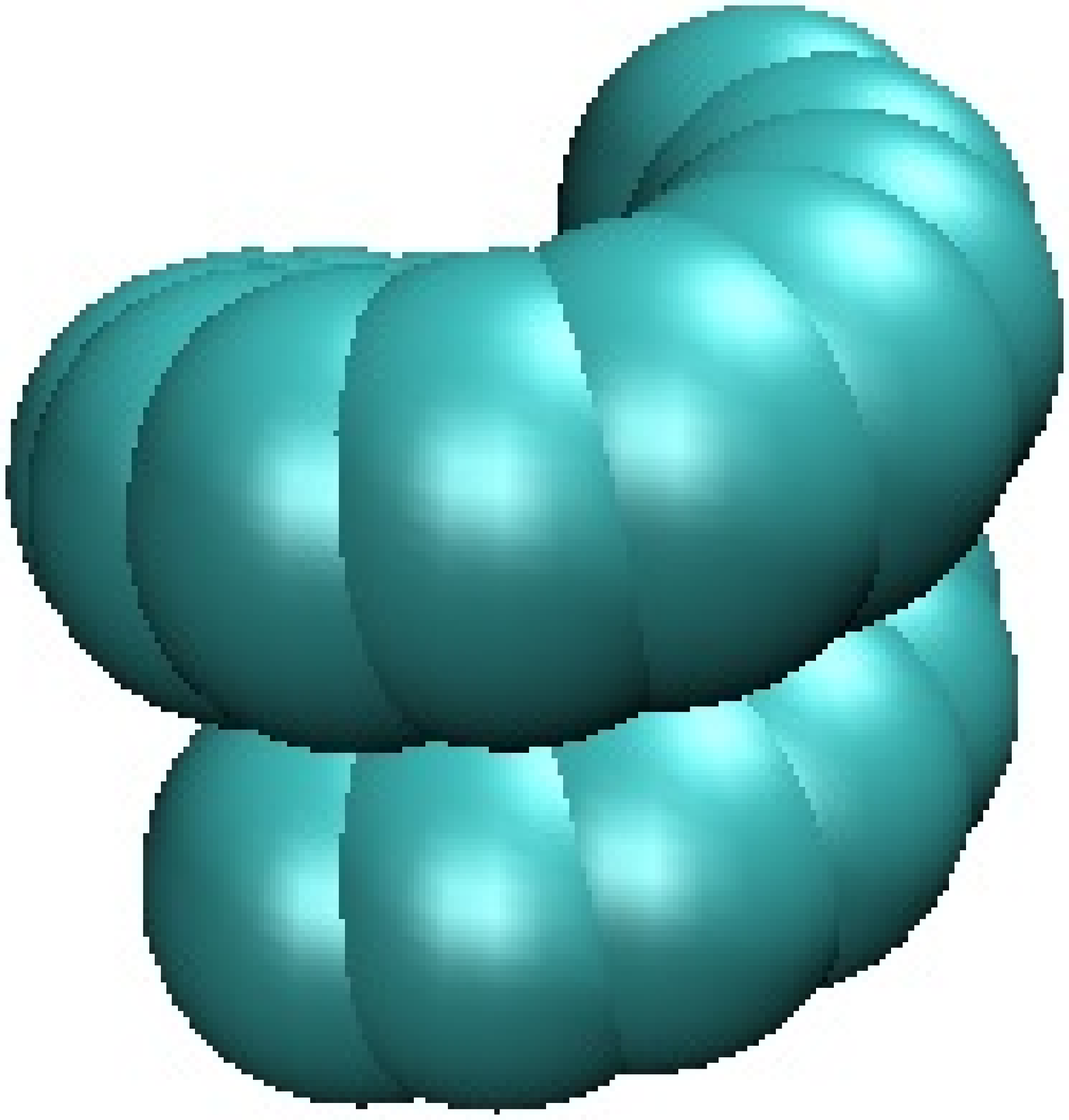,height=1.2in,width=1.2in}}
\label{f:helixcompare}
\caption{A comparison of the level-set (left) and SASA (right)
  description of helical polymer chain A with 32 atoms (above) and B
  with 22 atoms (below), respectively.}
% a model helical chain with 32 atoms.  Below: a model helical chain with 22 atoms.}
\end{figure}

From our MD simulations, we estimate solvation free energies of $\Delta
G\simeq 6$$k_BT$ and $\Delta G \simeq 7$$k_BT$ for the alkanes A and
B, respectively. The energies are positive and small and have the same
order of magnitude as the solvation energy of small
alkanes \cite{gunsteren,levy}. These relatively small free energies are
a well-known consequence of enthalpy-entropy compensation in the
solvation of nonpolar solutes \cite{gunsteren,levy}. It has been shown
that $\Delta G$ can be decomposed in solute-solvent enthalpy $\Delta
U_{\rm uv}$ and solute-solvent entropy $T\Delta S_{uv}$, while the
solvent-solvent enthalpy and entropy exactly cancel each other
\cite{gunsteren,levy}. For nonpolar solutes the enthalpic part is the
(ensemble-averaged) total solute-solvent LJ interaction which we
estimate by our MD simulations to be a large $\Delta U_{\rm
  uv}\simeq$-107$k_BT$ and -55$k_BT$ for the alkanes A and B,
respectively, indicating solute-solvent entropies of the same order of
magnitude.

In the level-set method, we can reproduce the total solvation free
energy for both alkanes within 5\% by using a Tolman length of
$\tau=1.3$\AA. In our implicit model, we can identify the
solute-solvent enthalpic part $\Delta U_{\rm uv}$ by the LJ term
$G_{\rm vdW}=:G_{\rm vdW}[v_{\rm min}]$ and the entropic part
$-T\Delta S_{\rm uv}$ by the interfacial term $G_{\rm sur}=:G_{\rm
  sur}[v_{\rm min}]$ in the free energy functional; we find large
values of $G_{\rm vdW}=-95$$k_BT$ and $G_{\rm sur}=101$$k_BT$ for
helix A, and $G_{\rm vdW}=-48$$k_BT$ and $G_{\rm sur}=55$$k_BT$ for
helix B, close to the values obtained by the MD simulations.  We note
that small variations of the Tolman length $\tau$ have
%We note that varying the Tolman length $\tau$ around 1\AA~has 
a negligible 
effect on the enthalpic part of the free energy while its influence on
the entropic (interfacial) part is noticeable. For example,  we find $G_{\rm
  sur}=63$$k_BT$ for helix B when using $\tau=1.1$\AA~ instead of
$G_{\rm sur}=55$$k_BT$ for $\tau=1.3$\AA.  As the total solvation free
energy is a difference of two large quantities, the interfacial part
of the free energy functional~(\ref{eq:grand}), in particular the
curvature correction, has to be reconsidered carefully.

In Fig.~3, we show the same two helical chains but using a color code
that indicates the value of mean curvature at each point of the
solute-solvent interface. The curvature varies between positive and
negative values, showing as well concave parts (blue). The highest
positive curvature (deep red) is roughly given by the inverse of the
length $\sigma$ of one LJ sphere. Note that the concave parts of the
loosely packed helix are within the helical core in contrast to the
convex outer parts.  This qualitatively different hydration of the
helix depending on the local (convex or concave) geometry is in line
with structural studies of water at protein-water
interfaces \cite{chothia}.

\begin{figure}[bhpt]
\centerline{\psfig{figure=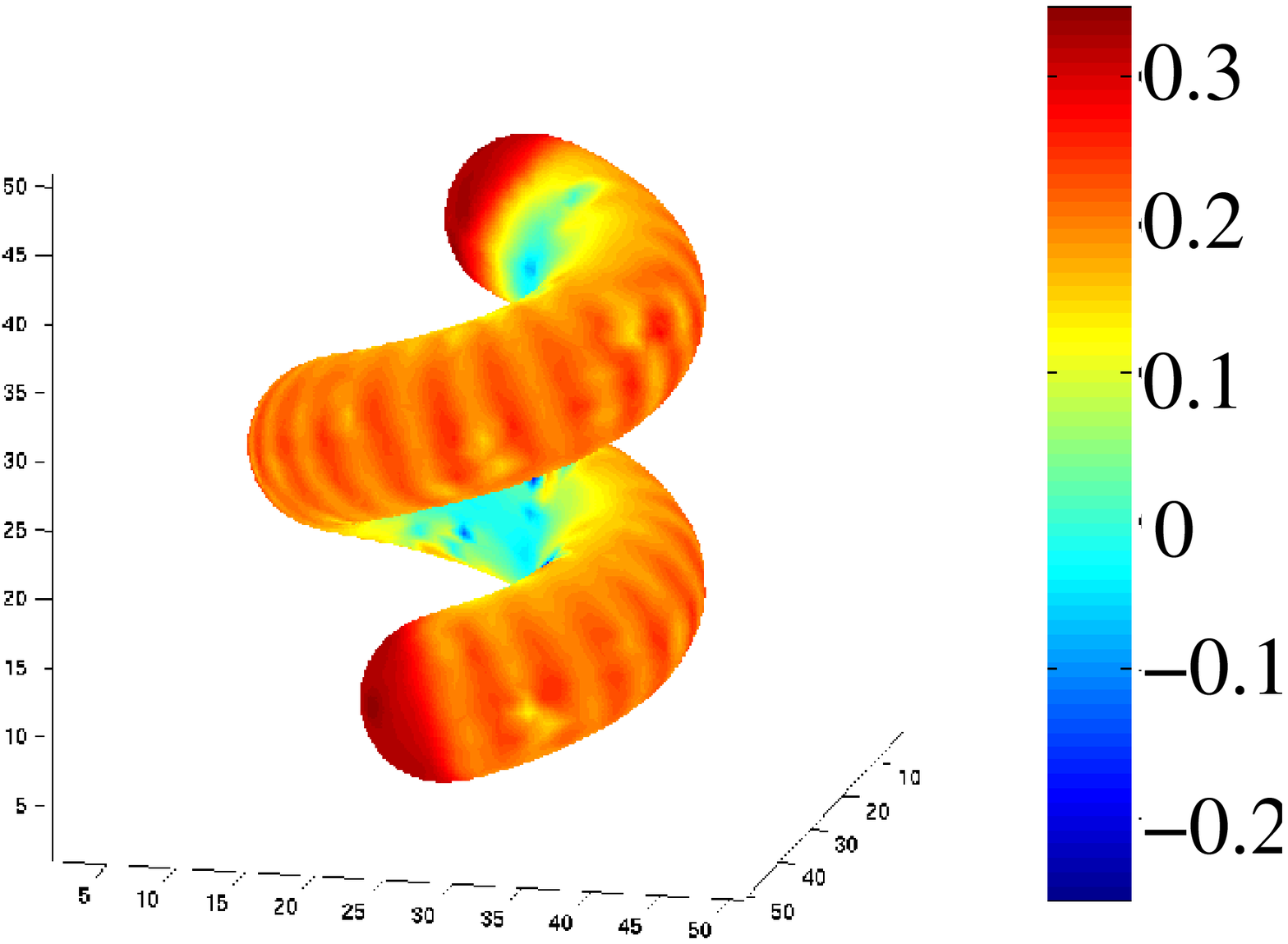,height=1.3in,width=1.6in}
\psfig{figure=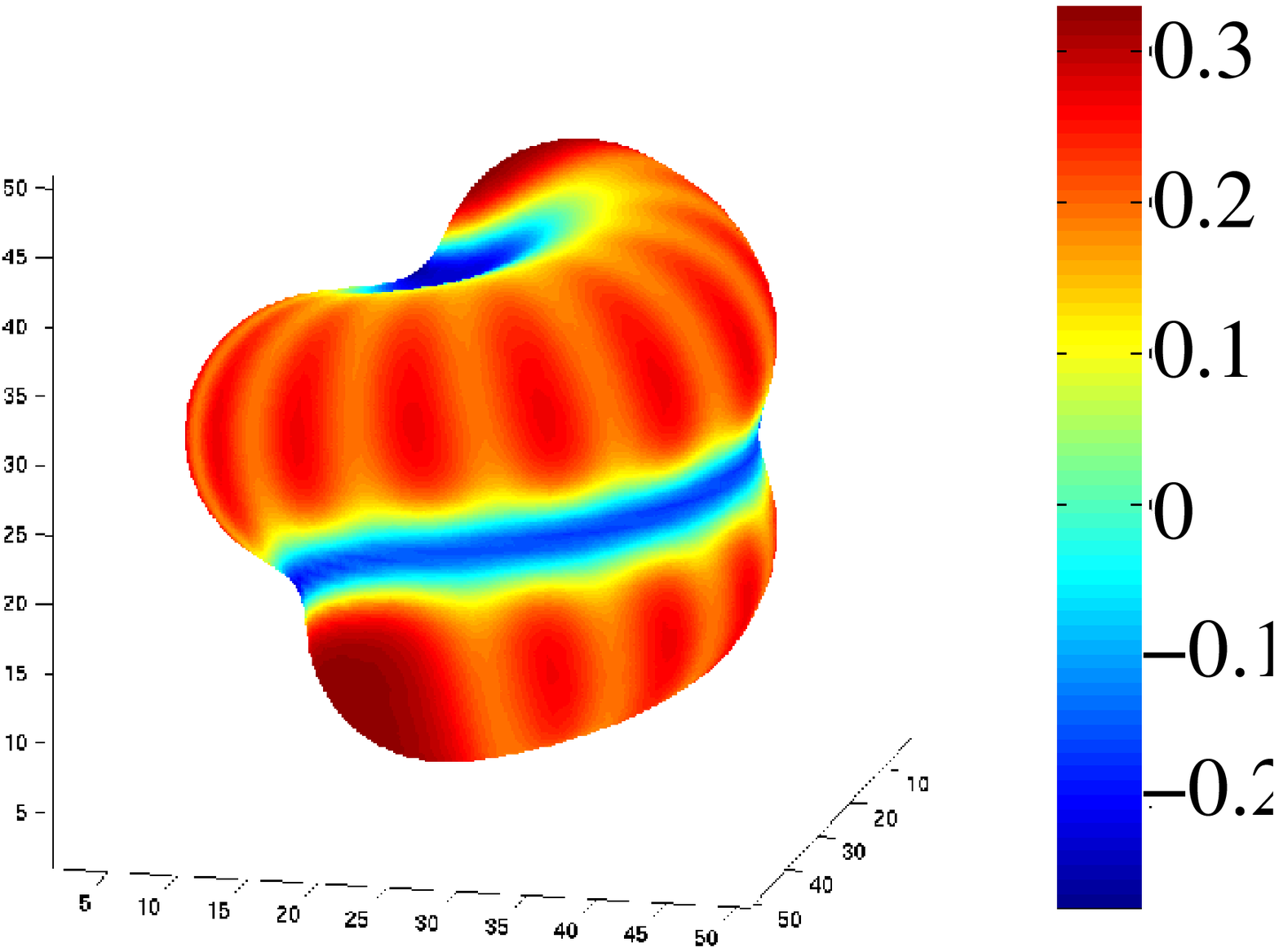,height=1.3in,width=1.6in}}
\label{f:HelixCurvature}
\caption{Level-set calculations of helical molecules. Color represents
  mean curvature. The units of mean curvature are in \AA$^{-1}$.}
\end{figure}

These examples show that complex shapes with varying interface
curvature, as typical in protein structures, can be efficiently
tackled by the variational implicit solvent model in conjunction with
the level-set method. The solute-solvent interface and the resulting
energies seem to be well described by our methods, in particular when
regarding the fact that the solvation free energy is a {\it
  difference} of large entropic and enthalpic contributions which
often leads to a large error in the calculation of the total solvation
free energy.

%%%%%%%%%%%%%%%%%%%%%%%%%%%%%%%%%%%%%%%%%%%%%%%%%%%%%%%%%%%%%%%%%%%%%%%%%%%%%%%%
\subsection{A single molecule of fullerenes $C_{60}$}  
\label{ss:C60}

Another interesting molecule which can be reasonably modeled as a
nonpolar entity is the C$_{60}$ molecule, the buckyball.  The
C-atom-water LJ parameters are taken from \cite{girifalco} and are
also shown in Tab.~I.  The experimental solvation free energy in water
is typically close to zero \cite{stukalin} in agreement with our MD
simulations which yield $\Delta G\simeq -1k_BT$. Previous more
empirical implicit solvent studies show that the large interfacial
energy penalty is more than compensated by the strong dispersion
attraction between the water and the tightly packed carbon shell
resulting in a small and negative total solvation free energy
\cite{stukalin}. A related unusual repulsive solvent-mediated
interaction between two C$_{60}$ molecules has been observed in MD
simulations \cite{bedrov}.

Our level-set result for the equilibrium interface is shown in Fig.~4,
featuring a smooth soccer-like surface. Using a Tolman length of
$\tau=1.3$\AA, we can reproduce the MD solvation energy within 5\%. A
separate analysis of the interfacial energy part and the dispersion
part shows that a large $\Delta G_{\rm vdW}=-49$$k_BT$ is gained from
dispersion attraction while $\Delta G_{\rm sur}=48$$k_BT$ surface
energy is paid. Our MD simulations confirm this cancellation of energy
contributions quantitatively by showing enthalpic and entropic
contributions of $\simeq -50$$k_BT$ and $\simeq 49$$k_BT$,
respectively.

\begin{figure}[bhpt]
\centerline{\psfig{figure=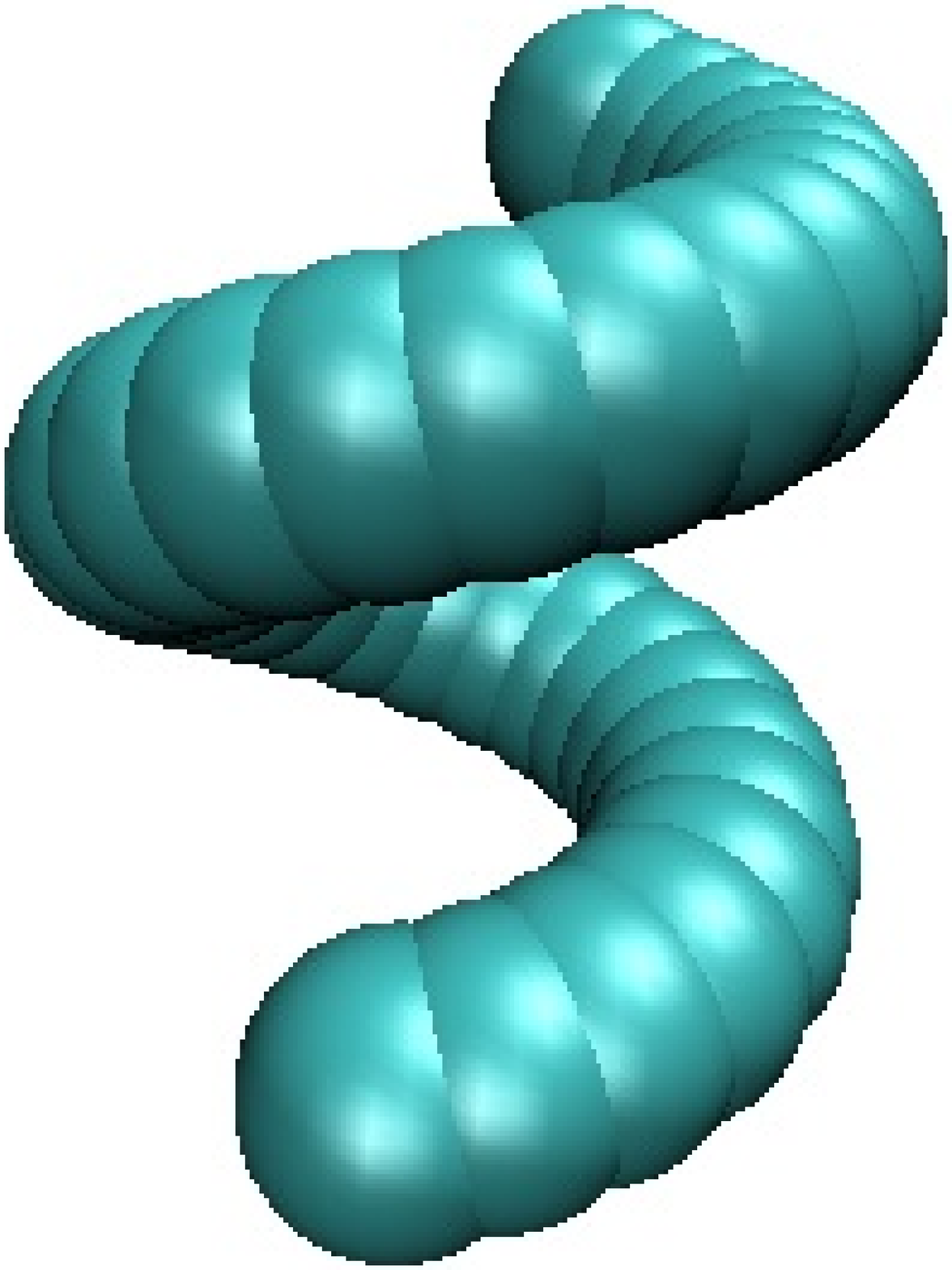,height=1.6in,width=1.6in}
\psfig{figure=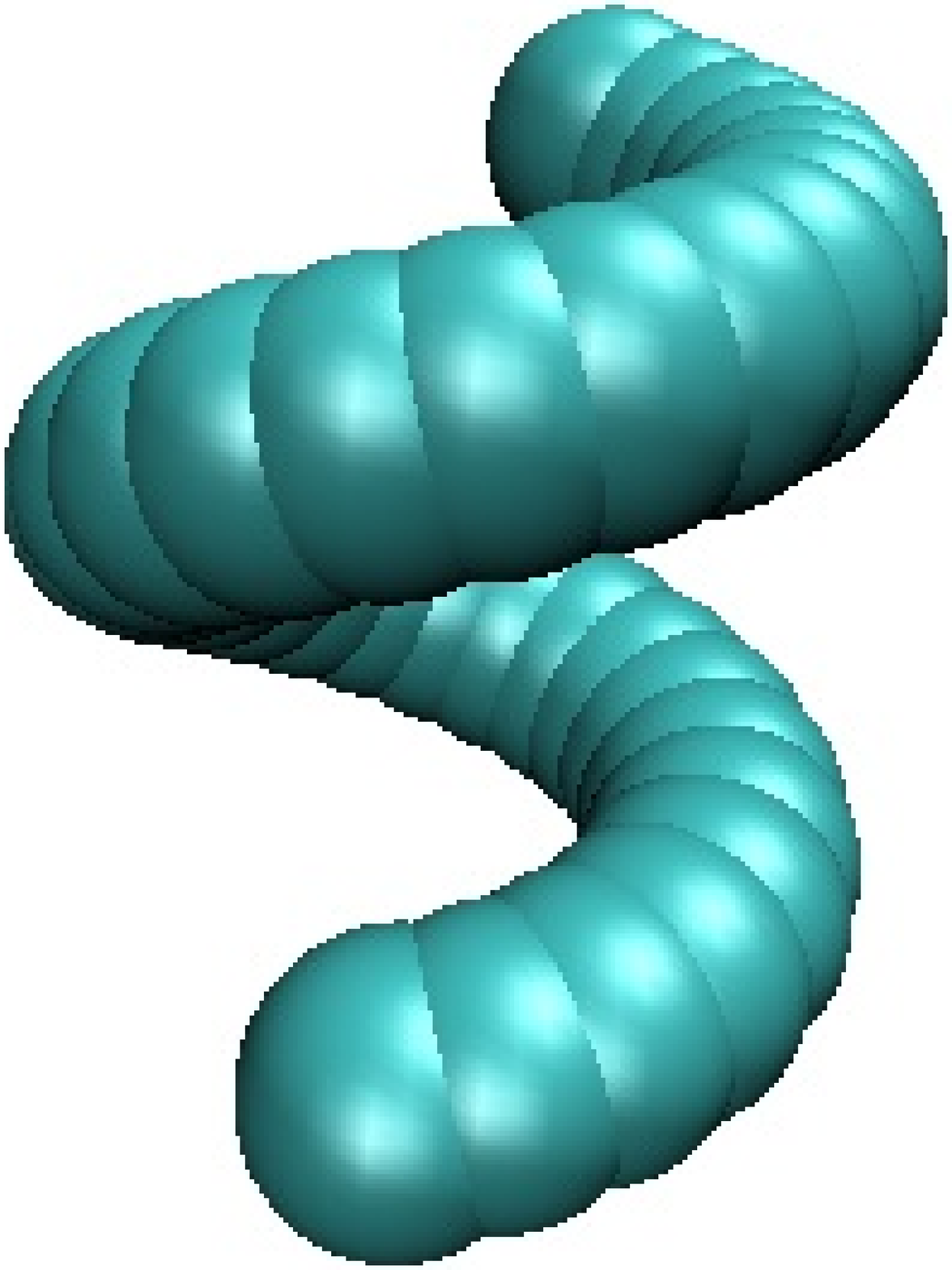,height=1.1in,width=1.1in}}
\label{f:soccer}
\caption{A comparison of the level-set (left) and a SASA type (right)
  description of a $C_{60}$ buckyball.}
\end{figure}

In Fig.~5, we show the same $C_{60}$ molecule obtained by our
level-set calculation with a color code that indicates the value of
mean curvature of the interface. In contrast to the helical molecules,
we find only convex curvatures varying from zero to roughly the
inverse of the LJ size of one C atom. The curvature distribution
displays the typical five and sixfold structure of the C$_{60}$
molecule.

\begin{figure}[bhpt]
\centerline{\psfig{figure=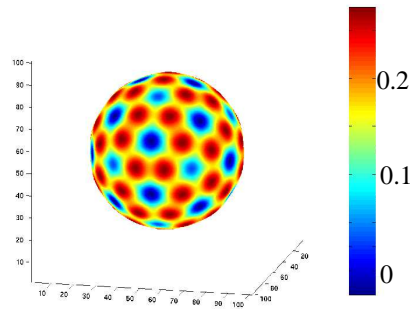,height=1.6in,width=2.1in}}
\label{f:soccercolor}
\caption{The level-set calculation of the $C_{60}$ molecule with
color indicating the value of mean curvature.
   The units of mean curvature are in \AA$^{-1}$.}
\end{figure}

As in the previous case of the two helical alkanes, this example
demonstrates that the subtle balance between interface (entropic) and
dispersion (enthalpic) terms, and thus the correct interface location
and curvature are crucial for an accurate description of solvation
free energies, and are well captured by our methods.

%%%%%%%%%%%%%%%%%%%%%%%%%%%%%%%%%%%%%%%%%%%%%%%%%%%%%%%%%%%%%%%%%%%%%%%%%%%%%%%%

\subsection{Two parallel nanometer-sized paraffin plates}
%\subsection{Dewetting and PMF between two nanometer-sized paraffin plates}
\label{ss:plates}

In our last example, we consider the strongly hydrophobic system of
two parallel paraffin plates as investigated in the explicit water MD
simulations by Koishi {\it et al.}~\cite{koishi}. Each plate consists
of $6 \times 6$ fixed CH$_2$-atoms with atom-water LJ parameters from
the OPLS-AA force field, see Tab.~1, and has a square length of
$\sim$3nm.  The two plates are placed in a center-to-center distance
$d$ and different separations are investigated.  Koishi {\it et
  al.}~observed a clear dewetting transition (vapor bubble or
``nanobubble'' formation) for distances $d \lesssim 15$\AA~accompanied
by a strong attractive interaction energy of the order of tens of
$k_BT$. The pmf is shown in Fig.~6 together with the solution of our
variational implicit solvent model and level-set snapshots of the
equilibrium interface at selected distances.

\begin{figure}[bhpt]
\centerline{\psfig{figure=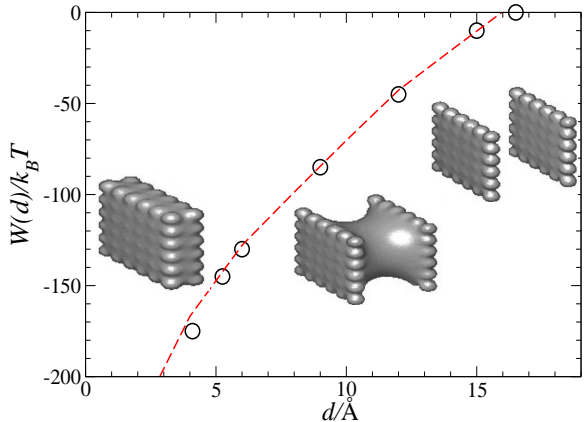,height=2.2in,width=3in}}
\label{f:plates}
\caption{Comparison of the pmf from level-set (circles) and MD
  calculations (dashed line) for the two-plate system.  Three
  level-set interfaces of the system are shown at their
  respective separations.}
\end{figure}

As can be seen in Fig.~6, we have almost perfect agreement with the MD
simulation results. We find that the curve hardly depends on the
particular choice of the Tolman length ($\tau=0.9$\AA~here).  The
reason is that the average radius of mean curvature for this
large-scale example is much bigger than the typical value of the
Tolman length and the theory basically becomes fit-parameter free. As
another important result, our variational implicit solvent model
captures the nanobubble formation, see the interface snapshots in
Fig.~6 for $d<15$\AA, a consequence of the energetic desire of the
system to minimize the total interface area.  For larger distances the
interface breaks, i.e., the bubble ruptures and the equilibrium
interface is given by two isolated plates having almost no mutual
interaction.

We note here that for distances $d<15$\AA~where bubble formation takes
place, the free energy of the system might have a second, local
minimum corresponding to the wetted case with two isolated plates \cite{LCW99}.
The existence of two local minima shows the possibility of hysteresis
in bubble formation as observed in the MD simulations \cite{koishi}.

%%%%%%%%%%%%%%%%%%%%%%%%%%%%%%%%%%%%%%%%%%%%%%%%%%%%%%%%%%%%%%%%%%%%%%%%%%%%%%%%
\section{Conclusions and outlook}
\label{s:conclusions}

We have developed a level-set method for numerically capturing stable
solute-solvent interfaces for {\it complex shaped} nonpolar molecules
in three dimensions based on a recently developed variational implicit
solvent model.  This method evolves an arbitrary initial surface to
decrease the solvation free energy until a stable equilibrium
solute-solvent interface is reached.
%So, this is an optimization process.  The key feature of our method is 
%that it can capture surface topological changes.  This enables us
%to capture the dewetting of an underlying system. 

In implementing our numerical method, we have developed a regularizing
technique to stabilize the surface evolution governed geometrically by
a combination of both the mean curvature and Gaussian curvature.  We
have additionally developed a stable, two-grid method for the calculation of
the total free energy.

We have applied our method to several nonpolar systems.  Our numerical
results show good agreement with MD simulations with reasonable
interface shapes and curvatures. In particular, due to the numerical
robustness and ability of handling topological changes, our method
captures the volume fusion/break and nanobubble formation/rupture in
the two xenon system and the strongly hydrophobic two-plate system,
respectively.

A comment is needed for the treatment of large concave curvatures
which can occur locally in microscopic systems. In the two Xe atom
example immediately before break-up, a singularity near the neck of
the two spheres develops and can artificially increase the energy. We
find that the energy is lowered if we renormalize the Tolman length
for very large mean curvatures.  Mathematically, it is known that even
the motion of surface solely by mean curvature can induce the
neck-pinch singularity that we see in our level-set calculations
\cite{AngenentVelazquez97}.  Therefore, it remains to be further
investigated how the singularity should be treated in the free energy
definition and level-set calculation.

We emphasize that our level-set implicit solvent calculations are one or two
orders of magnitude faster than explicit MD simulations.
Our method does require the solution of the level-set equation \reff{lse}
with the normal velocity \reff{expvn} in each time step. 
Compared with a SASA type approach, this is an ``extra'' work, but can be done efficiently.  
For instance, if we start with an initial surface that is close to an equilibrium one, 
our calculations can be much more efficient. 
With a good initial guess and a reasonable grid size, 
we need about $15$--$20$ minutes to run our code for the calculation 
of a helical polymer chain. 
When the level-set method is applied to
real dynamics calculations using continuum models, the relaxation of
the surface to a complete equilibrium is not necessary, and only a few
iterations are enough to update an interface. Therefore, in such a case, 
the level-set calculation can be compatible with a SASA type method 
in terms of efficiency. 

Our level-set method can be used for the force calculation of an underlying
solvation system.  In principle, forces are obtained by the gradient of the 
free energy with respect to some spatial coordinates. One such coordinate 
is the geometrical location of an equilibrium solute-solvent interface. 
Our artificial or optimization normal velocity is exactly the effective
force with respect to such a coordinate.  This replaces the calculation of surface
area in a SASA type method. Within the framework of level-set method, we
can also evaluate forces between the solute atoms that can be used as input 
to Brownian dynamics computer simulations. This is an important issue that
we are still investigating in details. 

Finally, let us comment on further developments which will be crucial
for a complete implicit solvation description of large biomolecules by
our method. We are currently developing a level-set method for
capturing numerically stable solute-solvent interfaces using the Gibbs
free energy \reff{eq:grand} that includes the electrostatic
contribution of the polar groups of the solutes. This method couples
the presented level-set method to a finite-difference based solver for
the Poisson-Boltzmann (PB) equation, a typical approach for the
implicit treatment of electrostatics in solvated molecular systems
\cite{DavisMcCammon90,SharpHonig90}.  In \cite{CCDLM07}, we derive the
variation of the free energy \reff{eq:grand} including electrostatics
on the PB level with respect to the change of the interface
$\Gamma$. This will be used to define the normal velocity $v_n$
similar to that in \reff{vnvn} but extended to local electrostatic
pressures.  Currently we are also developing fast and optimized level-set
methods for the handling of large systems that can have a few thousand
atoms in solutes and that can allow solute atoms to freely move around
in the optimization process. Another challenge is to extend the level-set
treatment to predict the correct {\it dynamics of evolution} of the
interface \cite{dzubiella:jcp:2007} and account for interface
fluctuations.
\\
%%%%%%%%%%%%%%%%%%%%%%%%%%%%%%%%%%%%%%%%%%%%%%%%%%%%%%%%%%%%%%%%%%%%%%%%%%%%%%%%

\noindent {\bf Acknowledgments.}
This work is partially supported by the NSF through grant DMS-0511766
(L.-T.C.) and DMS-0451466 (B.L.), by the DOE through grant
DE-FG02-05ER25707 (B.L.), and by a Sloan Fellowship (L.-T.C.).  J.D.\ thanks the Deutsche
Forschungsgemeinschaft (DFG) for support within the Emmy-Noether
Programme. Work in the McCammon group is supported in part by NSF,
NIH, HHMI, CTBP, NBCR, and Accelrys.  The authors thank Dr.~Jianwei
Che of Genomics Institute of the Novartis Research Foundation for
useful discussions and Dr.\ Dietmar Paschek for making the MD data in Fig.~1
available.

%%%%%%%%%%%%%%%%%%%%%%%%%%%%%%%%%%%%%%%%%%%%%%%%%%%%%%%%%%%%%%%%%%%%%%%%%%%%%%%%

\end{document}